\documentclass{webofc}
\usepackage[varg]{txfonts} 
\usepackage{xcolor}

\begin{document}

\title{A dispersive estimate of the $a_0(980)$ contribution to hadronic light-by-light scattering in $(g-2)_\mu$}

\author{\firstname{Oleksandra} \lastname{Deineka}\inst{1}\fnsep\thanks{\email{deineka@uni-mainz.de}} \and
        \firstname{Igor} \lastname{Danilkin}\inst{1}\fnsep \and
        \firstname{Marc} \lastname{Vanderhaeghen}\inst{1}\fnsep
}

\institute{Institut f\"ur Kernphysik \& PRISMA$^+$  Cluster of Excellence, Johannes Gutenberg Universit\"at,  D-55099 Mainz, Germany}

\abstract{
A dispersive implementation of the $a_0(980)$ resonance to $(g-2)_\mu$ requires the knowledge of the double-virtual $S$-wave $\gamma^*\gamma^*\to\pi\eta/ K\bar{K}_{I=1}$ amplitudes. To obtain these amplitudes we used a modified coupled-channel Muskhelishvili–Omn\`es formalism, with the input from the left-hand cuts and the hadronic Omn\`es function. The latter were obtained using a data-driven $N/D$ method in which the fits were performed to the different sets of experimental data on two-photon fusion processes with $\pi\eta$ and $K\bar{K}$ final states. This yields the preliminary dispersive estimate $a_\mu^{\text{HLbL}}[a_0(980)]_{\text{resc.}}=-0.46(2)\times 10^{-11}$.
}

\maketitle

\section{Introduction}\label{sec:intro}

The tension between the presently ultra-precise measurements of the anomalous magnetic moment of the muon $(g-2)_\mu$ and the theoretical calculations amounts to around $5.0\,\sigma$ difference \cite{Muong-2:2023cdq} when compared to the theoretical value from the 2020 White Paper \cite{Aoyama:2020ynm}. The source of the current theoretical error solely arises from contributions from hadronic vacuum polarization (HVP) and hadronic light-by-light scattering (HLbL). Apart from the pseudo-scalar pole contributions, further nontrivial contributions to HLbL arise from the two-particle intermediate states such as $\pi\pi$, $\pi\eta$, and $K\bar{K}$. Currently, only the contributions from the $\pi\pi_{I=0,2}$ and $K\bar{K}_{I=0}$ channels have been considered in a dispersive manner \cite{Colangelo:2017qdm, Danilkin:2021icn}. The isospin-0 part of this result can be understood as a model-independent implementation of the contribution from the $f_0(500)$ and $f_0(980)$ resonances. The contribution from the $a_0(980)$ resonance arises from the rescattering of the $\pi\eta/K\bar{K}_{I=1}$ states and necessitates knowledge of the double-virtual processes $\gamma^*\gamma^*\to\pi\eta/ K\bar{K}_{I=1}$. On the experimental side, currently, data is only available for the real photon case from the Belle Collaboration \cite{Belle:2009xpa, Belle:2013eck}. The measurement of the photon-fusion processes with a single tagged photon is a part of the two-photon physics program of the BESIII Collaboration \cite{Redmer:2018gah}. To describe the currently available data and provide theoretical predictions for the single- and double-virtual processes, we opt for the dispersive approach, which adheres to the fundamental properties of the $S$-matrix, namely, analyticity and coupled-channel unitarity.

\section{Formalism}\label{sec:formalism}

To compute the HLbL contribution of $a_0(980)$ to $(g-2)_\mu$, we adopt the formalism outlined in \cite{Colangelo:2017qdm}. This approach yields the following master formula:
\begin{equation}
a_\mu^{HLbL} = \frac{2\alpha^3}{3\pi^2}\int\limits_0^\infty dQ_1\int\limits_0^\infty dQ_2 \int\limits_{-1}^1 d\tau\sqrt{1-\tau^2}\,Q_1^3\,Q_2^3\sum_{i=1}^{12}T_i(Q_1,Q_2,Q_3)\,\bar{\Pi}_i(Q_1,Q_2,Q_3)\,,
\end{equation}
where $\bar{\Pi}_i$ are scalar functions containing the dynamics of the HLbL amplitude, $T_i$ denote known kernel functions, and $\tau$ is defined as $Q_3^2=Q_1^2+2Q_1Q_2\tau+Q_2^2$. For the $S$-wave, the only contributing scalar functions can be written as
\begin{align}\label{Eq:Pi3Pi9}
\bar{\Pi}^{J=0}_3&=\frac{1}{\pi}\int\limits_{s_{th}}^\infty ds' \frac{-2}{\lambda_{12}(s')(s'+Q_3^2)^2}\left(4s'\text{Im}\bar{h}_{++,++}^{(0)}(s')-(s'-Q_1^2+Q_2^2)(s'+Q_1^2-Q_2^2)\text{Im}\bar{h}^{(0)}_{00,++}(s')\right)\,,\nonumber\\
\bar{\Pi}^{J=0}_9&=\frac{1}{\pi}\int\limits_{s_{th}}^\infty ds' \frac{4}{\lambda_{12}(s')(s'+Q_3^2)^2}\left(2\,\text{Im}\bar{h}_{++,++}^{(0)}(s')-(s'+Q_1^2+Q_2^2)\,\text{Im}\bar{h}^{(0)}_{00,++}(s')\right)\,,
\end{align}
plus crossed versions. Here $\lambda_{12}(s) \equiv \lambda(s,Q_1^2,Q_2^2)$ is a K\"all\'en triangle function. 

Since $a_0(980)$ is known to have a dynamical coupled-channel $\pi\eta/K\bar{K}$ origin, the inclusion of $K\bar{K}$ intermediate states is necessary. In this case, the unitarity relation implies 
\begin{equation}
\text{Im}\bar{h}^{(0)}_{1,\lambda_1\lambda_2,\lambda_3\lambda_4}(s)=\bar{h}^{(0)}_{1,\lambda_1\lambda_2}(s)\,\rho_{\pi\eta}(s)\,\bar{h}^{(0)*}_{1,\lambda_3\lambda_4}(s)+\bar{k}^{(0)}_{1,\lambda_1\lambda_2}(s)\,\rho_{KK}(s)\,\bar{k}^{(0)*}_{1,\lambda_3\lambda_4}(s)\, ,
\end{equation}
where $\rho_{\pi\eta}(\rho_{KK})$ is the phase space factor of $\pi\eta\,(K\bar{K})$ system, and $\bar{h}^{(0)}_{1,\lambda\lambda'}\,(\bar{k}^{(0)}_{1,\lambda\lambda'})$ denotes the $I=1$, $J=0$ Born subtracted (e.g. $\bar{k}\equiv k-k^{\text{ Born}}$) partial-wave (p.w.) amplitude of the $\gamma^*(Q_1^2)\gamma^*(Q_2^2)\to \pi\eta\,(K\bar{K})$ process. 
These p.w. amplitudes contain kinematic constraints 
and therefore it is important to find a transformation to a new basis of amplitudes which can be used in a modified Muskhelishvili-Omn\`es (MO) method \cite{Garcia-Martin:2010kyn}.
For the $S$-wave, the amplitudes which are free from kinematic constraints can be written as \cite{Colangelo:2017qdm} \footnote{To maintain consistency with Eq.(\ref{Eq:Pi3Pi9}) we follow the conventions from \cite{Colangelo:2017qdm}, which slightly differ from those in \cite{Danilkin:2019opj}.}
\begin{equation}\label{Eq:constraints}
\bar{h}_{i=1,2}^{(0)} = \frac{\bar{h}_{++}^{(0)}\mp Q_1 Q_2\bar{h}_{00}^{(0)}}{s-s_\text{kin}^{(\mp)}}\, ,\quad s_\text{kin}^{(\pm)}\equiv - (Q_1\pm Q_2)^2\,,
\end{equation}
with $Q_i\equiv \sqrt{Q_i^2}$. In Eq.(\ref{Eq:constraints}) we omitted the isospin index for simplicity.
In the case of a single virtual or real photons, this constraint arises from the requirement of the soft-photon theorem. 
Similarly to $\gamma^*\gamma^*\to \pi\pi/K\bar{K}$ process \cite{Danilkin:2018qfn,Danilkin:2019opj}, the coupled-channel dispersion relation for the $\gamma^*\gamma^*\to \pi\eta/K\bar{K}$ process with $J=0,\,I=1$ can be written as follows
\begin{align}\label{eq:dr}
\left(\begin{array}{c}h^{(0)}_{i}(s)\\
k^{(0)}_{i}(s)\end{array}\right)=\left(\begin{array}{c}0\\
k^{(0), \text{ Born}}_{i}(s)\end{array}\right) +\Omega^{(0)}(s) \Bigg[-\int\limits_{s_{th}}^{\infty}\frac{ds'}{\pi}\,\frac{\text{Disc}(\Omega^{(0)}(s'))^{-1}}{s'-s}\left(\begin{array}{c}0\\ k^{(0),\text{ Born}}_{i}(s')\end{array}\right)
\Bigg]\,,
\end{align}
where only kaon-pole left-hand cut is currently taken into account. The generalization of the kaon-pole left-hand contribution $k_i^{(0),\text{Born}}$ to the case involving off-shell photons is achieved by the product of the scalar QED result with the electromagnetic kaon form factors \cite{Colangelo:2015ama}. The latter is parameterized using the VMD model. We have verified that within the $Q^2\lesssim 1$ GeV$^2$ range, which is crucial for the $a_\mu$ calculation, VMD is consistent with a simple monopole fit to the existing data and the dispersive estimation from \cite{Stamen:2022uqh}. 

To obtain the Omn\`es matrix $\Omega^{(0)}(s)$, which encodes the hadronic $\pi\eta/K\bar{K}$ rescattering effects, we utilize the coupled-channel dispersion relation for the partial wave amplitude. The latter is numerically solved using the $N/D$ ansatz \cite{Chew:1960iv}, with input from the left-hand cuts. When bound states or Castillejo-Dalitz-Dyson (CDD) poles are absent, the Omn\`es matrix is the inverse of the $D$-matrix. We parameterize the left-hand cuts in a model-independent manner, expressing them as an expansion in a suitably constructed conformal mapping variable \cite{Gasparyan:2010xz, Danilkin:2010xd}, which is chosen to map the left-hand cut plane onto the unit circle. In the absence of experimental $\pi\eta/K\bar{K}$ data, the coefficients of this conformal expansion can be estimated theoretically from $\chi$PT, as demonstrated in \cite{Danilkin:2011fz, Danilkin:2012ua, Danilkin:2017lyn}. However, for the $\pi\eta/ K\bar{K}$ system, it is necessary to rely on the slowly convergent $SU(3)$ $\chi$PT. Instead, we directly determine the unknown coefficients by fitting to $\gamma\gamma \to \pi\eta/K_S K_S$ data \cite{Belle:2009xpa, Belle:2013eck} and use $\chi$PT predictions only as additional constraints. Particularly, for the $\pi\eta\to K\bar{K}$ channel, we impose an Adler zero and ensure that the $\pi\eta\to K\bar{K}$ amplitude remains consistent with $\chi$PT at $s_{th}=(m_\pi+m_\eta)^2$. Furthermore, for the $\pi\eta\to\pi\eta$ channel, we employ the $\chi$PT scattering length as a constraint. In all cases, the NLO result with low-energy coefficients from \cite{Bijnens:2014lea} is considered as the central value, with an error range defined by the spread between LO and NLO results.

\section{Results and Outlook}\label{sec:results}

\begin{figure}
\centering
\includegraphics[width=0.49\textwidth]{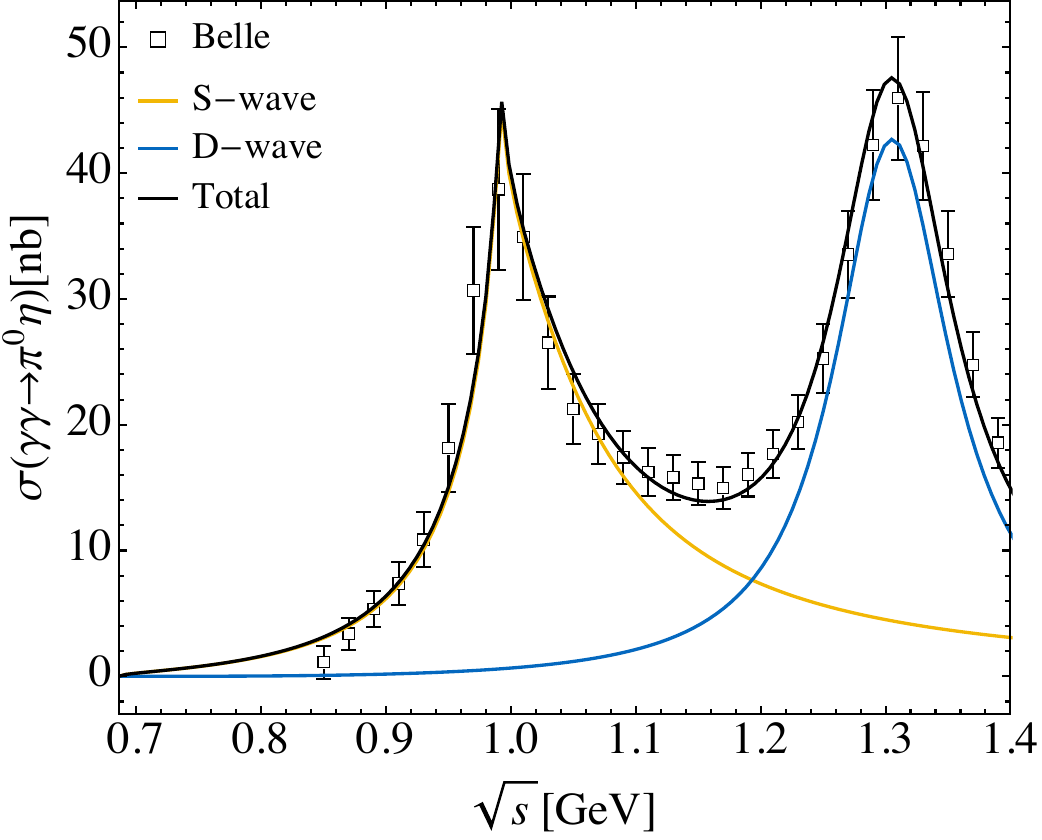}\,
\includegraphics[width=0.49\textwidth]{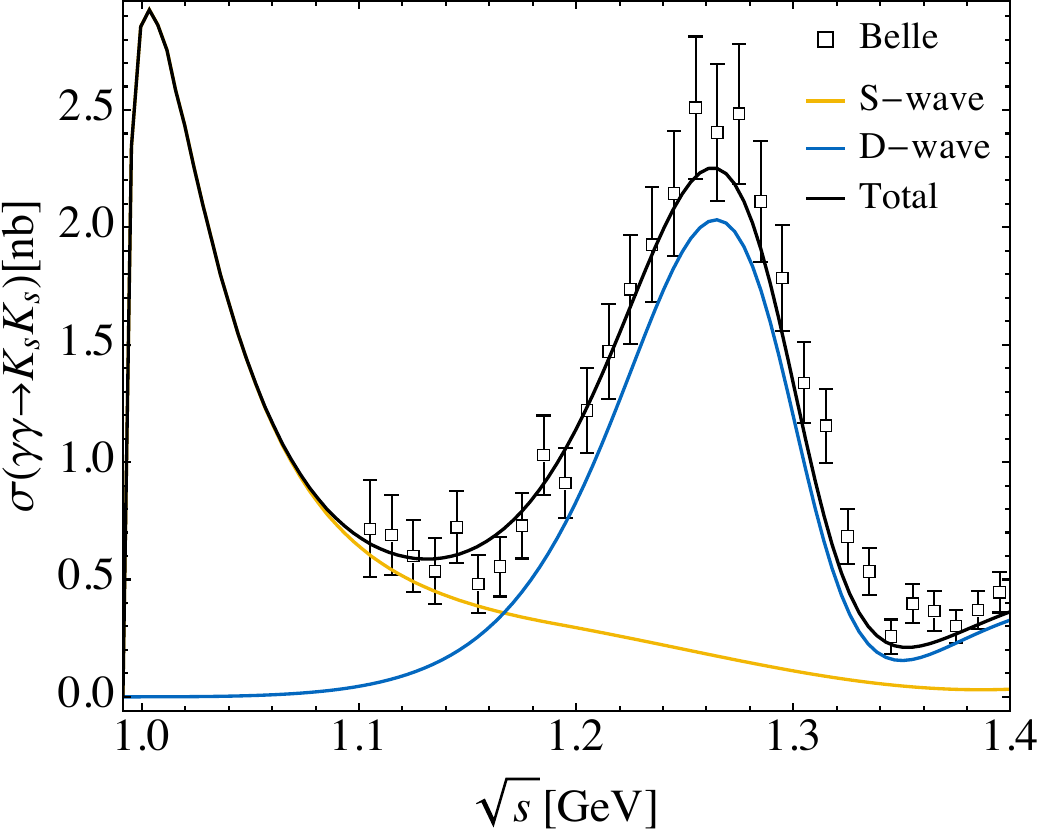}
\caption{Total cross-sections ($|\cos\theta|<0.8$) for the $\gamma\gamma\to\pi^0\eta$ (left) and $\gamma\gamma\to K_s K_s$ (right) processes compared to the fit results. The data are taken from \cite{Belle:2009xpa, Belle:2013eck}. }
\label{fig:cs}     
\end{figure}

To reconstruct the physical $\gamma\gamma\to K_S K_S$ cross section, the input for $I=0$, $S$-wave amplitude $k^{(0)}_{0,++}(s)$ is taken from the coupled-channel $\pi\pi/K\bar{K}_{I=0}$ analysis \cite{Danilkin:2020pak}. Since we are aiming to describe $\gamma\gamma \to \pi\eta/K_S K_S$ data in the region from threshold up to $1.4$ GeV, we also incorporate the $D$-wave resonances $f_2(1270)$ and $a_2(1320)$ using the Breit-Wigner parametrization, similar to the approach in \cite{Lu:2020qeo}. We find that with as few as $(2,2,2)$ $S$-wave parameters in $(11,12,22)$ channels ($1=\pi\eta, 2=K\bar{K}$) we obtain the fit with $\chi^2/\text{d.o.f.}=0.83$. The resulting total cross sections for $\gamma\gamma \to \pi\eta/K_S K_S$ processes are illustrated in Fig. \ref{fig:cs}. Through analytical continuation into the complex plane we find the pole on the Riemann sheet II, corresponding to the $a_0(980)$ resonance with $\sqrt{s_{a_0(980)}}=1.06 - i0.058$ GeV.

With the obtained $\gamma^*\gamma^*\to \pi\eta/ K\bar{K}$ amplitudes in hand, we can now proceed to calculate the $a_0(980)$ contribution to the HLbL in $(g-2)$. The preliminary result is
\begin{equation}
    a_\mu^{\text{HLbL}}[a_0(980)]_{\text{rescatering}}=-0.46(2)\times 10^{-11}\,,
\end{equation}
where the uncertainty currently covers only the sum-rule violation (reflecting the choice of the HLbL basis \cite{Colangelo:2017qdm}).
It is useful to compare the obtained dispersive result with the outcome from the narrow width approximation $   a_\mu^{\text{HLbL}}[a_0(980)]_{\text{NWA}}=-\left([0.3,0.6]^{+0.2}_{-0.1}\right)\times 10^{-11}$ \cite{Danilkin:2021icn},
where the range reflects the variation in the scale of transition form factor parametrisation taken from the quark model \cite{Schuler:1997yw}.

It is planned to further add new experimental data into the current analysis, in particular, $\gamma\gamma\to K^+K^-$ data from BESIII \cite{Kussner:2022dft}. In addition, the hadronic $\pi\eta/K\bar{K}$ rescattering will be further constrained by including the existing data for the $\phi\to\gamma\pi\eta$ \cite{KLOE:2009ehb} and $\eta' \to \pi\pi\eta$ \cite{BESIII:2017djm} decays.

\end{document}